# Thermostatistics of classical fields

Sergey A. Rashkovskiy

*Ishlinsky Institute for Problems in Mechanics of the Russian Academy of Sciences, Vernadskogo Ave., 101/1, Moscow, 119526, Russia*

*Tomsk State University, 36 Lenina Avenue, Tomsk, 634050, Russia*

*E-mail: rash@ipmnet.ru, Tel. +7 906 0318854*

**Abstract**
In this paper we develop the thermostatistics of the classical (continuous in space and time) fields. Assuming the thermodynamic equilibrium between the classical field and the thermal reservoir and the Gibbs statistics for the classical field, the probabilities of excitation of various modes of the classical field are determined. An artificial quantization of the classical field is considered, when the field energy is artificially divided into discrete portions - quanta with the same energy. The probability of excitation of a given number of quanta of a classical field is calculated. It is shown that if no restrictions are imposed on a classical field, then it obeys the Bose-Einstein statistics. If, however, an additional restriction is imposed on the amplitude of the classical field, it is described by a statistic analogous to the Fermi-Dirac statistics or the Gentile statistics, but does not coincide exactly with them.

## 1. Introduction

Traditionally, thermostatistics is constructed as a theory of systems consisting of discrete particles or quanta which is in equilibrium with a thermal reservoir.

It has been shown in [1-3] that many quantum phenomena can be described within the classical field theory if light and electron waves are considered as classical wave fields described by the Maxwell equations and Dirac equation, respectively, which are considered as a classical field equations.

From this point of view, it is interesting to construct a thermostatistics with respect to the classical (i.e., continuous in space and time) fields that are in equilibrium with the thermal reservoir.

The considered approach can also be applied to other classical fields, for example, electromagnetic, acoustic and hydrodynamic (turbulent) fields.

## 2. Gibbs statistics for classical fields

We consider the classical (i.e., continuous in space and time) field, which is in equilibrium with the thermal reservoir. For simplicity, we assume that the field is scalar and described in the general case by one scalar complex-valued function $\Phi(\mathbf{r}, t)$.

As usual, we consider the field localized in a certain domain.

We introduce an orthonormal system of basis functions $\{\varphi_s(\mathbf{r})\}, s = 0,1,2, ...$, satisfying the condition

$$\langle \varphi_s \varphi_r \rangle \equiv \int K(\mathbf{r}) \varphi_s^*(\mathbf{r}) \varphi_r(\mathbf{r}) dV = \delta_{sr} \tag{1}$$

where $K(\mathbf{r})$ is some function (the kernel of the system of basis functions); $V$ is the volume of the domain.

Such a system of basis functions is easy to introduce through the linear Hermitian operator $\hat{E}$ acting on spatial coordinates:

$$\hat{E} \varphi_s(\mathbf{r}) = \varepsilon_s \varphi_s(\mathbf{r}) \tag{2}$$

where $\varepsilon_s$ are some parameters. Thus, the basis functions $\varphi_s(x)$ are eigenfunctions of the operator $\hat{E}$, while the parameters $\varepsilon_s$ are its eigenvalues; we use the index $s$ simply as a mode number. Then the field $\Phi(\mathbf{r}, t)$ can be expanded with respect to the basis functions $\{\varphi_n(\mathbf{r})\}$:

$$\Phi(\mathbf{r}, t) = \sum_s g_s(t) \varphi_s(\mathbf{r}) \tag{3}$$

where $g_s(t)$ are the complex-valued amplitudes of the mode $s$, which are random functions of time.

Taking into account (1), one obtains

$$g_s(t) = \int K(\mathbf{r}) \varphi_s^*(\mathbf{r}) \Phi(\mathbf{r}, t) dV \tag{4}$$

Note that in the general case the classical field $\Phi(\mathbf{r}, t)$ can be nonlinear, i.e. it can satisfy a nonlinear field equation. In other words, in the general case, equation (2) is not a field equation for the classical field $\Phi(\mathbf{r}, t)$. However, it can always be expanded in a series (3) with respect to the eigenfunctions of the linear operator (2). Moreover, the different classical fields described by different field equations can be expanded with respect to the same system of basis functions $\{\varphi_s(\mathbf{r})\}$.

We introduce the integral characteristic of the field $\Phi(\mathbf{r}, t)$

$$E = \langle \Phi \hat{E} \Phi \rangle \equiv \int K(\mathbf{r}) \Phi^*(\mathbf{r}, t) \hat{E} \Phi(\mathbf{r}, t) dV \tag{5}$$

Taking into account certain arbitrariness in the choice of a system of basis functions, we choose them in such a way that the parameter $E$ is equal to the energy of the field.

We consider the classical fields $\Phi(\mathbf{r}, t)$ for which this integral converges; thus the field energy $E$ is finite.

Taking into account (1) - (3), one obtains

$$E = \sum_s \varepsilon_s |g_s|^2 \qquad (6)$$

This expression can be interpreted as follows: the field energy is composed of the energy of all eigenmodes, while the energy of each mode is equal to $\varepsilon_s |g_s|^2$. The parameters $\varepsilon_s$ are uniquely determined by the choice of the system of basis functions $\varphi_s(\mathbf{r})$ (see Eq. (2)). The random amplitudes $|g_s|^2$ of different modes of the classical field are not subject to any restrictions; they vary continuously and theoretically they can take any values in the range $[0, \infty)$.

Thus, at each instant of time the field $\Phi(\mathbf{r}, t)$ is completely characterized by a set of all complex-valued amplitudes $\{g_s\}$.

We introduce the phase space for the classical field (3).

Writing complex-valued amplitudes in the form

$$g_s = q_s + ip_s \qquad (7)$$

where $q_s$ and $p_s$ are the real-valued functions of time, we see that at each instant of time the field $\Phi(\mathbf{r}, t)$ is completely characterized by a set of real-valued parameters $\{p, q\}$, where $p = (\dots, p_1, p_2, \dots)$; $q = (\dots, q_1, q, \dots)$.

We introduce an infinite-dimensional phase space $\{p, q\}$ whose coordinates are the parameters $q_s$ and $p_s$ for all $s$. Obviously, this phase space is Hilbert space if the complex-valued amplitudes $g_s$ are used instead of the real-valued parameters $q_s$ and $p_s$.

The state of the field $\Phi(\mathbf{r}, t)$ at each instant of time can be represented by a point in the phase space $\{p, q\}$. If the field $\Phi(\mathbf{r}, t)$ changes over time, the representative point moves in the phase space, describing a certain phase trajectory. Thus, the evolution (change) of the field over time is depicted, as usual, by a phase trajectory in the phase space $\{p, q\}$.

We denote the elementary volume of the phase space $\{p, q\}$ by $dpdq = \cdots dp_1 dq_1 dp_2 dq_2 \dots dp_s dq_s \dots$.

Then the probability that the classical field $\Phi(\mathbf{r}, t)$ is in a state close to the state with amplitudes of the excited eigenmodes $g_n$ is

$$dP(\dots, g_1, g_2, \dots) = f(\dots, g_1, g_2, \dots) dpdq \qquad (8)$$

where $f(\dots, g_1, g_2, \dots)$ is the probability density of such a state of the classical field.

By analogy with the way it is done in classical thermostatistics, it is easy to show that the stationary (equilibrium) distribution for the classical field $\Phi(\mathbf{r}, t)$ must be Gibbsian:

$$f(\dots, g_1, g_2, \dots) = Z^{-1} \exp\left(-\frac{E}{kT}\right) \qquad (9)$$

where $Z$ is the partition function, which is defined as

$$Z = \int \exp\left(-\frac{E}{kT}\right) dpdq \qquad (10)$$

Taking (6) into account, one obtains

$$f(\ldots, g_1, g_2, \ldots) = Z^{-1} \exp\left(-\frac{\sum_s \varepsilon_s |g_s|^2}{kT}\right) \tag{11}$$

Taking into account that the complex-valued amplitudes $g_s$ enter the distribution function (11) in the form $|g_s|^2 = q_s^2 + p_s^2$, it is convenient to go to the polar coordinates $r_s$ and $\varphi_s$, such that $g_s = r_s \exp(i\varphi_s)$. Here $r_s \in [0, \infty)$; $\varphi_s \in [0, 2\pi]$.

In this case, the probability (8) takes the form

$$dP(\ldots, g_1, g_2, \ldots) = Z^{-1} \exp\left(-\frac{\sum_s \varepsilon_s r_s^2}{kT}\right) \ldots r_1 dr_1 d\varphi_1 r_2 dr_2 d\varphi_2 \ldots \tag{12}$$

In the special case when only one eigenmode $s$ of the field $\Phi(\mathbf{r}, t)$ is excited, one obtains

$$dP_s(g_s) = Z_s^{-1} \exp\left(-\frac{\varepsilon_s r_s^2}{kT}\right) r_s dr_s d\varphi_s \tag{13}$$

where

$$Z_s = \int_0^{2\pi} \int_0^{\infty} \exp\left(-\frac{\varepsilon_s r_s^2}{kT}\right) r_s dr_s d\varphi_s \tag{14}$$

As a result, one obtains

$$Z_s = \frac{\pi kT}{\varepsilon_s} \tag{15}$$

Obviously, if several eigenmodes of the field $\Phi(\mathbf{r}, t)$ are excited simultaneously, then the distribution function has the form (11), where

$$Z = \prod_s Z_s \tag{16}$$

Here the product is taken over all excited modes.

Thus, the thermostatistics of the classical fields under consideration are completely determined.

## 3. Artificial quantization of classical fields

We first consider the classical field $\Phi(\mathbf{r}, t)$, in which only one eigenmodes $s$ is excited. In this case, according to (6), the field energy is equal to

$$E_s = \varepsilon_s |g_s|^2 \tag{17}$$

We note that for the classical field $\Phi(\mathbf{r}, t)$, the energy changes continuously and can take any values from the admissible range. This is due to the fact that the amplitude $|g_s|^2$ of such a field can vary continuously.

At the same time, the relation (17) can be interpreted as follows.

We divide the entire possible continuous energy range into identical intervals (cells) of length $\varepsilon_s$. Then the field energy (17) falls into one of these discrete intervals (cells).

We can write the continuous parameter $|g_s|^2$ in the form

$$|g_s|^2 = N + \theta \tag{18}$$

where $N = 0,1,2,...$ is the integer part of the number $|g_s|^2$, $\theta$ is the fractional part of the number $|g_s|^2$, thus $0 < \theta < 1$.

Next, we are interested not in the exact value of energy (17), but only in the number of interval (cell) in which the energy (17) falls. Such a procedure can be called quantization of the classical field, however, it is obvious that in this case such quantization is artificial and in no way related to the existence of actual quanta of the field energy, since the field under consideration is classical, by definition,. Naturally, with this approach (artificial quantization), we lose some information about the field. As a result, there is an uncertainty in the exact value of the field energy, which can take any values within the length of the cell $\varepsilon_s$.

However, in some cases, such an approach can be justified by the fact that such a "quantized" field is easier to describe and analyze.

With this approach, the "quantized" energy of all fields in which actual energy falls into the same interval is considered the same.

We will first assume that the field $\Phi(\mathbf{r},t)$ is not in equilibrium with the thermal reservoir. The energy of the field (17) is determined by external impacts, which are considered random but not related to the temperature.

In this case, the energy (17) can take any values from a continuous interval.

We consider all possible states of the field that correspond to the same value of the integer parameter $N$, but to different values of the parameter $\theta$.

If the parameter $\theta$ can take any random values from the range $(0,1)$ with equal probability, then the mean value of this parameter is equal to

$$\langle \theta \rangle = \frac{1}{2} \tag{19}$$

Then the mean value of the energy of the classical field, corresponding to the same value of the integer parameter $N$ (that is, the conditional mean value, for a fixed value of $N$), is equal to

$$\langle E_s \rangle_N = \varepsilon_s \left( N + \frac{1}{2} \right) \tag{20}$$

Formally, relation (20) coincides with the expression for the energy spectrum of a quantum linear oscillator, although in this case we are dealing with a continuous classical field $\Phi(\mathbf{r},t)$.

Taking into account the formal similarity of relation (20) with the expression for the energy of a quantum linear oscillator, we will say that a classical field $\Phi(\mathbf{r},t)$ satisfying condition

$$N \leq |g_s|^2 < N + 1, \tag{21}$$

is in the energy state with $N$ "quanta" (contains $N$ "indivisible energy quanta").

Then expression (20) describes the mean value of the energy of the classical field $\Phi(\mathbf{r},t)$ for a fixed number of quanta $N$.

Let us consider what happens if the classical field $\Phi(\mathbf{r},t)$ is in equilibrium with the thermal reservoir.

Let us find the probability that the classical field $\Phi(\mathbf{r},t)$ is in the energy state with $N$ "quanta". This means that the parameter $r_s^2 = |g_s|^2$ is in the range $[N, N+1]$. In this case, using the distribution (13), (15), one obtains

$$P_s(N) = \frac{\varepsilon_s}{kT} \int_N^{N+1} \exp\left(-\frac{\varepsilon_s x}{kT}\right) dx \tag{22}$$

or

$$P_s(N) = \left[1 - \exp\left(-\frac{\varepsilon_s}{kT}\right)\right] \exp\left(-\frac{\varepsilon_s}{kT} N\right) \tag{23}$$

Obviously,

$$\sum_{N=0}^{\infty} P_s(N) = 1 \tag{24}$$

The mean "occupation number" ("population") of the eigenmode $s$ of the classical field $\Phi(\mathbf{r},t)$, which is in equilibrium with the thermal reservoir,

$$\langle N \rangle = \sum_{N=0}^{\infty} N P_s(N) \tag{25}$$

Calculating (25), using (23), one obtains

$$\langle N \rangle = \frac{1}{\exp\left(\frac{\varepsilon_s}{kT}\right) - 1} \tag{26}$$

Thus, we obtained the Bose-Einstein statistics for the classical field $\Phi(\mathbf{r},t)$, in the case when we are not interested in a detailed description of the field, but only its energy up to an artificial partition into discrete "quanta".

The mean energy of the mode $s$ of the classical field for any its populations

$$\langle E_s \rangle = \frac{\varepsilon_s}{kT} \int_0^{\infty} \varepsilon_s x \exp\left(-\frac{\varepsilon_s x}{kT}\right) dx$$

Calculating the integral, one obtains

$$\langle E_s \rangle = kT \tag{27}$$

Using (13), (15), and (23), we can calculate the mean energy of the mode $s$ of the classical field which is in the state with $N$ "quanta" (conditional mean value for fixed $N$):

$$\langle E_s \rangle_N = \frac{\varepsilon_s}{kT P_s(N)} \int_N^{N+1} \varepsilon_s x \exp\left(-\frac{\varepsilon_s x}{kT}\right) dx$$

Integrating, one obtains

$$\langle E_s \rangle_N = \varepsilon_s \left(N + \frac{kT}{\varepsilon_s} - \frac{1}{\exp\left(\frac{\varepsilon_s}{kT}\right) - 1}\right) \tag{28}$$

It is interesting to compare expression (28) for the mean energy of the "quantized" mode $s$ of the classical field, which is in equilibrium with the thermal reservoir, with expression (20) for the mean energy of the same "quantized" field, but in the free state. It can be seen that they have the same structure:

$$\langle E_s \rangle_N = \varepsilon_s (N + \langle \theta_s \rangle) \tag{29}$$

but with different values of the parameter $\langle\theta_s\rangle$: for a free field, according to (19), it is constant and equal to 0.5, while for a field which is in equilibrium with a thermal reservoir,

$$\langle\theta_s\rangle = \frac{kT}{\varepsilon_s} - \frac{1}{\exp\left(\frac{\varepsilon_s}{kT}\right)-1} \qquad (30)$$

The dependence (30) is shown in Fig. 1. It is seen that for a field which is in equilibrium with a thermal reservoir, the value of the parameter $\langle\theta_s\rangle$ is always less than the value (19) for the free field, and decreases with increasing $\frac{\varepsilon_s}{kT}$. And only for $\frac{\varepsilon_s}{kT} \to 0$ (for example, at a very high temperature $kT \gg \varepsilon_s$), the parameter $\langle\theta_s\rangle$ for a field which is in equilibrium with the thermal reservoir approaches the value (19) for the free field. On the contrary, as the parameter $\frac{\varepsilon_s}{kT}$ increases, the value of the parameter $\langle\theta_s\rangle$ decreases rapidly, and for $kT \ll \varepsilon_s$ the energy of the classical field which is in equilibrium with the thermal reservoir can approximately be written in the form

$$\langle E_s\rangle_N \approx \varepsilon_s N \qquad (31)$$

In this case, with sufficient accuracy, we can say that the classical field $\Phi(\mathbf{r},t)$ consists of an integer number of "indivisible quanta" having an energy $\varepsilon_s$. Moreover, the larger $\frac{\varepsilon_s}{kT}$, the more accurate is this statement.

Thus, we see that as the temperature is lowered, the classical field $\Phi(\mathbf{r},t)$ manifests more and more "quantum" properties: it seems that the field consists of an integer number of indivisible "quanta" having the energy $\varepsilon_s$.

It follows from (29) that in a state with $N = 0$ the field energy

$$\langle E_s\rangle_0 = \varepsilon_s\langle\theta_s\rangle \qquad (32)$$

Such a state of the classical field could be called a vacuum state in the sense that the field does not contain integer "quanta". In this state, the classical field has the lowest possible (at a given temperature) mean energy. The energy of the classical field in this state can be called zero-point energy. However, it should be borne in mind that in this case neither the vacuum state of the field nor the zero-point energy contain any special physical meaning, but are only conditional names used to denote a certain state of the classical field when energy of the mode $s$ of the field is in the range $[0, \varepsilon_s)$.

In the case when a classical field $\Phi(\mathbf{r},t)$ which is in equilibrium with a thermal reservoir, has simultaneously several excited eigenmodes, the mean field energy is equal to

$$\langle E\rangle = \sum_s \varepsilon_s(\langle N_s\rangle + \langle\theta_s\rangle) \qquad (33)$$

where $\langle N_s\rangle$ is determined by the relation (26), while $\langle\theta_s\rangle$ is determined by the relation (30); summation is over all excited eigenmodes. Obviously, in this case there is no divergence of energy when summing over an infinite number of eigenmodes, since the energy of the classical field remains finite and is simply redistributed between modes.

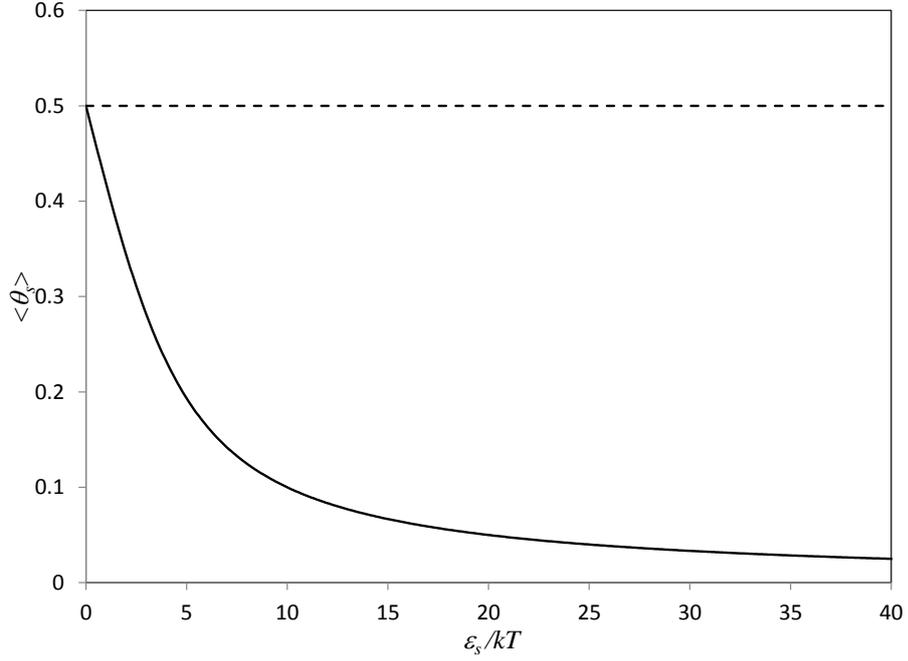

Fig. 1. Dependence (30) of the parameter $\langle \theta_s \rangle$ on the parameter $\frac{\varepsilon_s}{kT}$ (solid line). The dashed line shows the value (19) for the free field.

## 4. Fermionic classical fields

The classical field $\Phi(\mathbf{r}, t)$ considered above can be conditionally called a bosonic field, since discrete "quanta" obtained by artificial partition of the continuous field energy into equal intervals (cells), obey the Bose-Einstein statistics (26).

This is due to the fact that no restrictions were placed on the amplitude of such a field, and it was assumed that the modulus of amplitude of each excited mode can take any positive values, up to infinity.

We pose the question: can there be a classical field whose "quanta" obey the Fermi-Dirac statistics?

Consider the classical field $\Phi(\mathbf{r}, t)$, which has an additional normalization

$$\int K(\mathbf{r}) \Phi^* \Phi dV = n \qquad (34)$$

where $n=1,2,3,\ldots$ ($n=0$ means no field).

Substituting (3) into (34), one obtains

$$\sum_s |g_s|^2 = n \qquad (35)$$

This means that the field energy (5), (6) can change only in such a way that the condition (35) is satisfied.

This imposes additional restrictions on the amplitudes of the excitation of different modes.

We note that in the case of a bosonic field, the field energy (17) was directly quantized, while in the case of a fermionic field (34), the field amplitude itself (for example, its electric charge of the field [2, 3]) is quantized.

In this case, taking into account condition (35) the distribution function (11) for the classical field $\Phi(\mathbf{r}, t)$ has the form

$$f(\ldots, g_1, g_2, \ldots) = Z^{-1} \exp\left(-\frac{\sum_s \varepsilon_s |g_s|^2}{kT}\right) \delta(\sum_s |g_s|^2 - n) \tag{36}$$

where $\delta$ is the delta function; the partition function $Z$ is determined by the expression (10).

To calculate the partition function, we assume that there are a finite number of excited modes $M$, and then we can go to the limit $M \to \infty$.

Passing to the polar coordinates $r_s$ and $\varphi_s$, and, denoting $x_s = r_s^2$, taking into account the properties of the delta function, one obtains

$$Z_M = \pi^M \exp(-\beta_M n) \int_{\sum_{s=1}^{M-1} x_s \leq n} \exp(-\sum_{s=1}^{M-1} (\beta_s - \beta_M) x_s) \, dx_1 dx_2 \ldots dx_{M-1} \tag{37}$$

where $\beta_s = \frac{\varepsilon_s}{kT}$.

The integral in (37) is well known [4]:

$$\int_{\substack{\sum_{s=1}^{M} x_s \leq n \\ x_1 \geq 0, x_2 \geq 0, \ldots, x_M \geq 0}} \exp(\sum_{s=1}^{M} a_s x_s) \, dx_1 dx_2 \ldots dx_M =$$

$$n^M \int_{\substack{\sum_{s=1}^{M} y_s \leq 1 \\ y_1 \geq 0, y_2 \geq 0, \ldots, y_M \geq 0}} \exp(\sum_{s=1}^{M} a_s n y_s) \, dy_1 dy_2 \ldots dy_M = \sum_{s=1}^{M} \frac{\exp(a_s n) - 1}{a_s \varphi'(a_s)}$$

where

$$\varphi(u) = (u - a_1)(u - a_2) \ldots (u - a_M)$$

Obviously,

$$\varphi'(a_s) = \sum_{m=1}^{M} \prod_{\substack{r=1 \\ r \neq m}}^{M} (a_s - a_r)$$

Then for the integral (37), one obtains

$$\int_{\substack{\sum_{s=1}^{M-1} x_s \leq n \\ x_1 \geq 0, x_2 \geq 0, \ldots, x_M \geq 0}} \exp(-\sum_{s=1}^{M-1} (\beta_s - \beta_M) x_s) \, dx_1 dx_2 \ldots dx_{M-1} = \sum_{s=1}^{M-1} \frac{\exp(-(\beta_s - \beta_M) n) - 1}{\tau_s}$$

where

$$\tau_s = \sum_{m=1}^{M-1} \prod_{\substack{r=1 \\ r \neq m}}^{M} (\beta_r - \beta_s) \tag{38}$$

Finally, one obtains

$$Z_M = \pi^M \sum_{s=1}^{M-1} \frac{\exp(-\beta_s n) - \exp(-\beta_M n)}{\tau_s} \tag{39}$$

Using (39), one can calculate the mean occupation number of the mode $s$, which by definition is equal to

$$\langle N_s \rangle = \int |g_s|^2 f(\ldots, g_1, g_2, \ldots) dp dq \qquad (40)$$

By analogy with (37), one obtains

$$\langle N_s \rangle_M = Z^{-1} \pi^M \exp(-\beta_M n) \int_{\sum_{s=1}^{M-1} x_s \leq n} x_s \exp\left(-\sum_{s=1}^{M-1} (\beta_s - \beta_M) x_s\right) dx_1 dx_2 \ldots dx_{M-1} \qquad (41)$$

Obviously,

$$\langle N_s \rangle_M = -Z^{-1} \pi^M \exp(-\beta_M n) \frac{d}{d\beta_s} \int_{\sum_{s=1}^{M-1} x_s \leq n} \exp\left(-\sum_{s=1}^{M-1} (\beta_s - \beta_M) x_s\right) dx_1 dx_2 \ldots dx_{M-1} \qquad (42)$$

Taking into account (37) and (35), one obtains

$$\langle N_s \rangle_M = -Z_M^{-1} \frac{dZ_M}{d\beta_s}, s \neq M \qquad (43)$$

$$\langle N_M \rangle_M = n - \sum_{s=1}^{M-1} \langle N_s \rangle_M \qquad (44)$$

As an example, using (39), we can calculate $\langle N_s \rangle_M$ for several values of $M$.

For $M = 2$, one obtains

$$Z_2 = \pi^2 \frac{\exp(-\beta_1 n) - \exp(-\beta_2 n)}{(\beta_2 - \beta_1)} \qquad (45)$$

$$\langle N_1 \rangle_2 = \frac{1 - [1 + n\beta_{12}] \exp(-n\beta_{12})}{[1 - \exp(-n\beta_{12})] \beta_{12}} \qquad (46)$$

$$\langle N_2 \rangle_2 = n - \langle N_1 \rangle_2 \qquad (47)$$

where $\beta_{12} = \beta_1 - \beta_2 = \frac{\varepsilon_1 - \varepsilon_2}{kT}$.

The Fermi-Dirac distribution in the notation adopted has the form

$$\langle N_1 \rangle_{FD} = \frac{1}{\exp(\beta_{12}) + 1} \qquad (48)$$

The dependences (46) and (48) are shown in Fig. 2. It can be seen that for $n = 1$ the distribution (46) is qualitatively similar to the Fermi-Dirac distribution, however, it differs quantitatively from the Fermi-Dirac distribution.

It is interesting to note that if we use $4\beta_{12}$ instead of $\beta_{12}$ in distribution (46), then distribution (46) will be very close to the Fermi-Dirac distribution.

For other values of the parameter $n$, the distribution (46) is qualitatively similar to the function $n\langle N_1 \rangle_{FD}$, or, more accurately, to the Gentile distribution [5,6], though quantitatively also differs from it.

For $M = 3$, one obtains

$$Z_3 = \pi^3 \frac{\exp(-\beta_1 n)}{(\beta_2 - \beta_1)(\beta_3 - \beta_1)} + \pi^3 \frac{\exp(-\beta_2 n)}{(\beta_1 - \beta_2)(\beta_3 - \beta_2)} + \pi^3 \frac{\exp(-\beta_3 n)}{(\beta_1 - \beta_3)(\beta_2 - \beta_3)} \qquad (49)$$

In the general case, the distributions (43) for $M = 3$ depend on the relation between the energies of the eigenmodes $\varepsilon_1, \varepsilon_2$ and $\varepsilon_3$.

Let us consider a special case

$$\beta_3 - \beta_2 = \beta_2 - \beta_1 = \beta_{21} \tag{50}$$

In this case $\beta_3 - \beta_1 = 2\beta_{21}$ and

$$\langle N_1 \rangle_3 = -\frac{(3-2\beta_{21}n)-4\exp(-\beta_{21}n)+\exp(-2\beta_{21}n)}{2\beta_{21}(1-\exp(-n\beta_{21}))^2} \tag{51}$$

$$\langle N_1 \rangle_3 = \frac{1-2\beta_{21}n\exp(-\beta_{21}n)-\exp(-2\beta_{21}n)}{\beta_{21}(1-\exp(-\beta_{21}n))^2} \tag{52}$$

$$\langle N_3 \rangle_3 = n - \langle N_1 \rangle_3 - \langle N_2 \rangle_3 \tag{53}$$

Dependencies (51) and (52) for different values of the parameter $n$ are shown in Fig. 3 and 4. We see that for $n = 1$ the distribution (51) differs more from the Fermi-Dirac distribution than the distribution (46). However, it should be borne in mind that they were calculated for the special case (50).

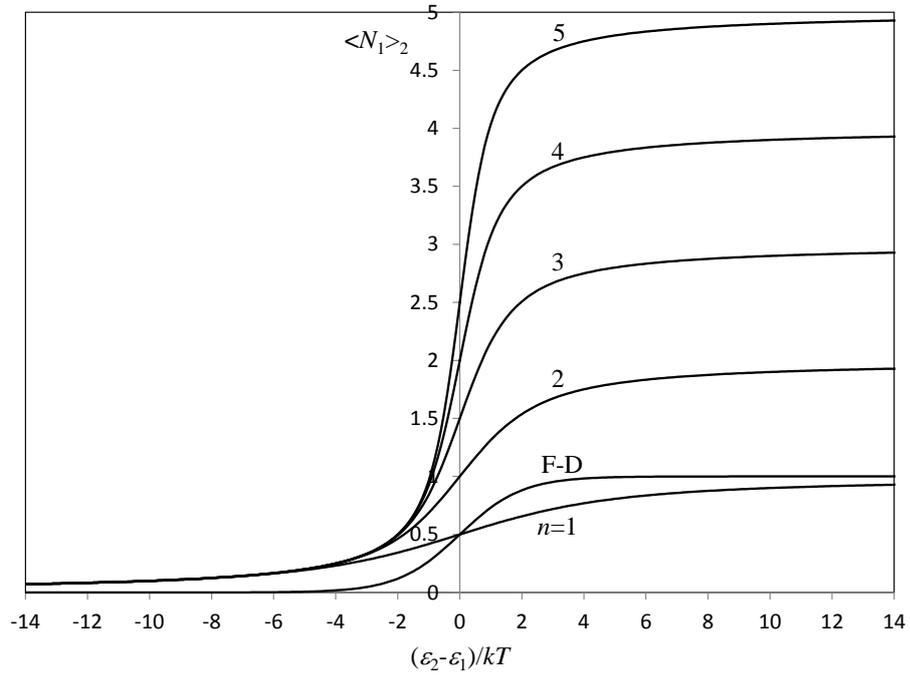

Fig. 2. Dependence of the mean occupation number of the mode $s = 1$ for $M = 2$ on the parameter $\frac{\varepsilon_2-\varepsilon_1}{kT}$ for different values of $n$. For comparison, the Fermi-Dirac distribution (48) is shown.

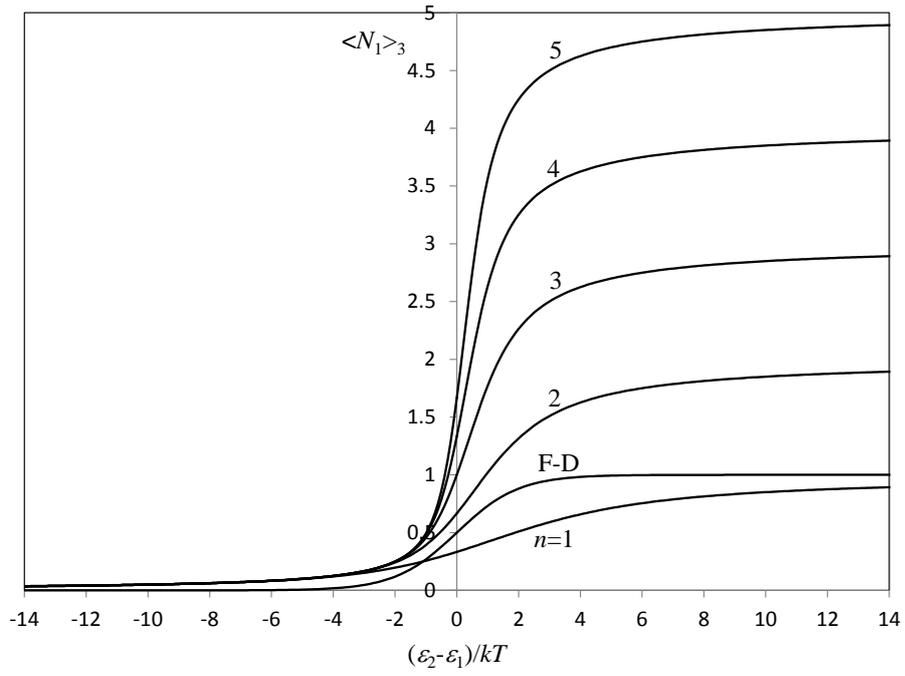

Fig. 3. Dependence of the mean occupation number of the mode $s = 1$ for $M = 3$ on the parameter $\frac{\varepsilon_2 - \varepsilon_1}{kT}$ for different values of $n$ for the case (50). For comparison, the Fermi-Dirac distribution (48) is shown.

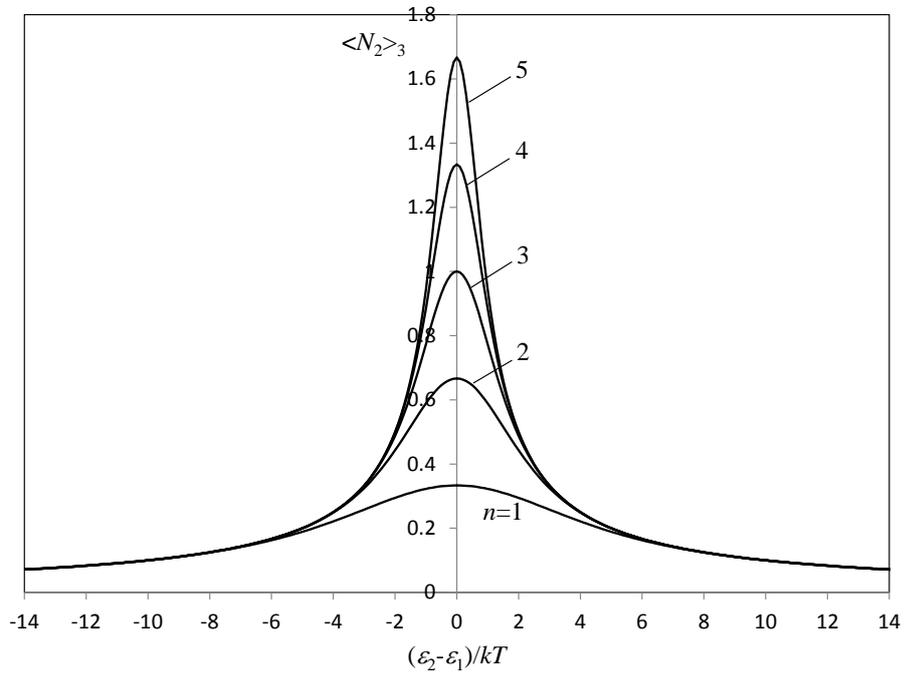

Fig. 4. Dependence of the mean occupation number of the mode $s = 2$ for $M = 3$ on the parameter $\frac{\varepsilon_2 - \varepsilon_1}{kT}$ for different values of $n$ for the case (50).

## 6. Concluding remarks

The above analysis shows that for a classical (i.e., continuous in space and time) field which is in equilibrium with a thermal reservoir, it is possible to construct in a natural way a thermostatistics, based on the Gibbs distribution.

In some cases, a detailed description of the classical field can be replaced by a simplified description using artificial quantization of the field energy. With the artificial quantization of the classical field, we again (as usual in thermostatistics) deal with a discrete system of "quanta", for which, naturally, without any additional assumptions and postulates, we obtain results well known from quantum statistics. In particular, it turned out that classical "bosonic" fields at very low temperatures have pseudo-quantum properties: it seems that they consist of a discrete number of indivisible quanta.

For the "quanta" of the classical "bosonic" field which is in equilibrium with the thermal reservoir, the Bose-Einstein statistics are obtained in a natural way, while for the "fermionic" field which is in equilibrium with the thermal reservoir, the statistics close to the Fermi-Dirac and Gentile statistics was obtained. This indicates that many of the results of quantum statistics can apparently be obtained within the framework of classical field theory, when the electromagnetic field and the electronic field are considered as the classical fields [1-3].

Using the results obtained we can conclude that the classical fields can also be described in some approximation using a second quantization approach. This issue is expected to be considered in the subsequent paper.

In this paper we considered the scalar field $\Phi(\mathbf{r}, t)$. Obviously, the results obtained can easily be generalized to other types of fields: vector, tensor, spinor, bispinor, and so on.

## Acknowledgments
This work was done on the theme of the State Task No. AAAA-A17-117021310385-6. Funding was provided in part by the Tomsk State University competitiveness improvement program.